\documentclass[aps,prb,superscriptaddress,twocolumn,tightenlines]{revtex4-1}
\usepackage{amssymb}
\usepackage{graphicx}
\usepackage{natbib}
\usepackage{epstopdf}
\usepackage{dcolumn}
\usepackage{bm}
\usepackage{makeidx}
\usepackage{enumerate}
\usepackage {graphicx}
\usepackage {amsmath}
\usepackage {amsfonts}
\usepackage {amssymb}
\usepackage {mathrsfs}
\usepackage {bbm}
\usepackage{subfigure}
\begin{document}

\title{Quasiclassical Eilenberger theory of the topological proximity effect in a superconducting nanowire}
\author{Valentin Stanev}
\affiliation{Condensed Matter Theory Center, Department of Physics,
University of Maryland, College Park, MD 20742-4111, USA}

\author{Victor Galitski}
\affiliation{Joint Quantum Institute and Condensed Matter Theory Center, Department of Physics,
University of Maryland, College Park, MD 20742-4111, USA}
\date{\today }

\begin{abstract}
We use the quasiclassical Eilenberger theory to study the topological superconducting proximity effects between a segment of a nanowire with a 
$p$-wave order parameter and a metallic segment. This model faithfully represents key qualitative features of an experimental setup, where only
a part of a nanowire is in immediate contact with a bulk superconductor, inducing topological superconductivity. It is shown that the Eilenberger equations 
represent a viable alternative to the Bogoliubov-de Gennes theory of the topological superconducting heterostructures and provide a much simpler quantitative description
of some observables. For our setup, we obtain exact analytical solutions for the quasiclassical Green's functions and the density of states as a 
function of position and energy. The correlations induced  by the boundary involve terms associated
with both $p$-wave  and odd-frequency pairing, which  are intertwined and contribute to observables on an equal footing. We recover the signatures of the
standard Majorana mode near the end of the superconducting segment, but find no such localized mode induced in the metallic segment. Instead,
the zero-bias feature is spread out across the entire metallic part in accordance with the previous works. In shorter wires, the Majorana mode and delocalized peak 
split up away from zero energy.  For long metallic segments, non-topological Andreev bound states appear and 
eventually merge together, giving rise to a gapless superconductor. 
\end{abstract}
\maketitle

\section{Introduction}
In recent years, there has been an explosion of interest in various solid-state architectures to realize Majorana fermions\cite{Reviews}. Many ingenious setups have been proposed
theoretically\cite{Fu,JDSau,Lutchyn, Alicea,Oreg,Wimmer,Nazarov}, and several experimental groups have reported promising results in superconducting heterostructures\cite{Mourik, Deng, Das}. However, it is not  just any Majorana fermion that the 
condensed matter physicists are interested in, but rather a localized Majorana, captured within a defect, or near a system's edge\cite{Kitaev}.


Most theoretical proposals involve superconducting proximity systems in which a common $s$-wave superconductor induces pairing in an electronic system, where only one electron species -- {\it i.e.}, one component of the Kramers doublet originating from the spin -- plays a role. The resulting description of the latter system  is often assumed to be equivalent to a mean-field theory of a topological $p$-wave superconductor, and therefore it becomes a Majorana host.  Such a Bogoliubov-de Gennes (BdG) description is justified if the entire mesoscopic system is in contact or coated with a massive superconductor, as indeed has been assumed in theoretical proposals. However, what actually is studied experimentally is not always that, because experiment often involves only a part of the potential host in direct contact with the superconductor, while the other part, actually being probed for localized Majoranas, is not in such contact\cite{setup}.

The ``leakage'' of superconductivity, topological or otherwise, into such a metallic system is {\em not} described by the mean-field BdG equations with a pairing term in the interior. The latter may only originate from a pairing interaction term in the underlying Hamiltonian, and  can not possibly appear in the absence of interactions. To be precise, BdG equations can still be written, but the proximity effect appears there only as a boundary condition describing Andreev reflection processes. If a nanowire lies on top of a bulk superconductor, this boundary condition can be rewritten as a $\delta$-function term in the BdG equation and integrated across the unimportant transverse direction, hence formally recovering a mean-field BdG theory with a minigap (see, Ref. \onlinecite{VG1} where such a calculation is demonstrated for a topological superconductor). However, if a wire extends away from the superconductor, a boundary Andreev term remains such and can not be replaced with a minigap away from the edges. 

This geometry leads to a different description, which in fact has been discussed extensively in the literature in the context of both conventional proximity systems and superconductor-ferromagnet junctions. The theoretical techniques used there typically involve either Eilenberger equations\cite{Eilenberger} for ballistic systems or Usadel equations\cite{Usadel} to describe the limit of strong disorder. One of hallmark results of these studies is the prediction of odd-frequency pairing, which allows spin-triplet s-wave correlations to appear\cite{EfetovRw} (with the Pauli exclusion principle enforced via the odd frequency-dependence of the Green functions, representing non-trivial retardation effects\cite{OFP}). This prediction enjoys strong experimental support in ferromagnetic proximity structures\cite{OFP_Exp}, which are quite relevant to the systems discussed in the context of Majorana physics. Several recent papers \cite{Tanaka1,Tanaka2,Valls,Linder,TanakaR} emphasized that the appearance of odd-frequency pairing is ubiquitous and actually unavoidable in many such setups. Furthermore, this pairing plays a special role in the emergence of topological edge states\cite{Asano}. However, the current literature does not seem to contain a clear message in relation to the actual Majorana experiments and shed light on the observable features that would differentiate it from the na{\"\i}ve mean-field theory.  


In this paper, we focus on these questions in the simplest possible model, which we believe retains all key ingredients relevant to actual experiment, and explore the nature of topological proximity effect into an extended one-dimensional ballistic metallic channel, with an eye on the structure of the density of states and in particular zero-energy modes there. We choose to use Eilenberger equations to do so. The upside of this treatment is that it is much simpler than Gor'kov or BdG equations, allows straightforward analytical solution, and provides the ingredients for a density of states (DOS) calculation right away. The downside of Eilenberger equations, which represent a ``lower level'' theory than BdG equations, is that certain information have been integrated out and interpretation of the nature of fermionic excitations contributing to the DOS is not straightforward. 

Despite this limitation, we proceed to derive a number of interesting results for the DOS in various geometries involving nanowires with superconducting $p$-wave and metallic segments. We demonstrate the presence of an $s$-wave, odd-frequency correlations in the Green's function, induced at the boundary between the segments. We study the role of these correlations and the appearance of in-gap resonances in the DOS of the metallic part. We interpret the zero-energy peak as a signature of a Majorana-type mode, which, however, is distributed over the entire normal region. This is consistent with several earlier analytic\cite{JD1,Klinovaja} and  numerical\cite{JD2,Chevallier} calculations, grounded in  BdG equations. We also study the effects of a single impurity inside the normal part. Finally, we consider a finite wire -- in this case the inter-edge mixing shifts the localized Majorana peak and delocalized Majorana-type mode to a finite energy.

The structure of the paper is as follows. The formalism we use is introduced in Sec.~\ref{EE}. Sec.~\ref{IW} applies this formalism to the simple case of a wire with infinite superconducting and metallic segments and derives a counter-intuitive result that despite the presence of strong superconducting correlations in the initially normal part, the density of states there is indistinguishable from that in a normal metal. In Sec.~\ref{IWFN}, we consider the case of a wire with an infinite superconducting part, but a finite  normal segment, which gives rise to the modification of metallic DOS and the appearance of a delocalized zero-energy peak. We further explore this geometry in Sec.~\ref{WI}, and study the effect of a single impurity in the normal part on the density of states. Sec.~\ref{FW} deals with the case of a wire with finite superconducting and metallic segments and demonstrates that the interplay of the localized Majorana mode in the former and delocalized mode in the latter gives rise to energy splitting of these features away from zero energy. We summarize our results in Sec.~\ref{DC}.

\section{Eilenberger equations}
\label{EE}

 We consider a one-dimensional system (we refer to it as ``nanowire'' throughout the text) of spinless fermions, which are in a $p$-wave superconducting state in a part of the wire ($x<0$). To describe it, we start with the two-particle Green's functions, which can be combined into a $2\times2$ matrix $\hat{G}_{\omega}({\bf p}_1, {\bf p}_2)$ in the particle-hole (Nambu) space. This matrix has both normal (particle-hole) and anomalous (particle-particle) parts, and is a function of the Matsubara frequency $\omega$ and the momenta of the two particles. This function contains information about all length-scales, including the shortest length scales of order Fermi wave-length $\sim \hbar/p_F$. This information is sometimes excessive to describe observables in experiments, where the relevant length-scales (e.g., the size of a normal tunnelling contact used to probe DOS) is larger than the Fermi wave-length. In such situations, it is advantageous to introduce a simplified  Green's function $ \hat{g}_{\omega}({\bf R}, {\bf p}_F/p_F) $, which can be obtained from $\hat{G}$ by integrating over the energies close to the Fermi surface. This quasiclassical Green's function, $ \hat{g}_{\omega}$, only depends on the center-of-mass coordinate for the pair ${\bf R}$  and the direction of the electrons' relative momentum on the Fermi surface ${\bf p}_F/p_F$. This matrix is the basis for the quasiclassical methods in superconductivity\cite{QGF} and it obeys the Eilenberger equation\cite{Eilenberger} (essentially a simplified version of the Gor'kov's equations). 

Since we consider a one-dimensional system,  $\hat{g}_{\omega}$ becomes especially simple. Its center-of-mass coordinate reduces to $x$ -- the one-dimensional coordinate along the wire, and the direction of the momentum on the Fermi surface, which we denote as $\zeta={p_x}/p_F$ in what follows, now can take on only two values: $\zeta=+1/-1$ for right/left moving fermions. The Eilenberger equation in the ballistic limit takes the form:
\begin{eqnarray}
\label{Eilenberger1}
 \zeta v_{F}\partial_x\hat{g}_{\omega} = [ \omega \hat{\tau_3}, \hat{g}_{\omega}] +i [\hat{\Delta}, \hat{g}_{\omega}], 
\end{eqnarray}  
where $\hat{\tau}_i$ denote the Pauli matrices in the Nambu space and $\left[ \cdot, \cdot \right]$ denotes a commutator of two-by-two matrices. We want to study a $p$-wave order parameter, and write $\hat{\Delta}$ in the general form $\hat{\Delta}\equiv \Delta_\zeta (x) \hat{\tau}_2$, with $\Delta_\zeta(x) = \zeta \Delta_0(x)$. We reiterate that in Eq.~(\ref{Eilenberger1}), the Green's function is a function of frequency ($\omega$), position ($x$), and $\zeta$. In addition to the Eilenberger equation~(\ref{Eilenberger1}), the Green's function $ \hat{g} $ also has to obey the normalization condition $\hat{g}^2=\hat{1}$, where $\hat{1}$ is the $2\times2$ unit matrix.

We can decompose $\hat{g}$ as:
\begin{eqnarray}
 \hat{g}_{\omega}(x, \zeta) = g_{1}(x, \omega, \zeta)\hat{ \tau}_1 + g_{2}(x, \omega, \zeta) i \hat{\tau}_2+  g_3(x, \omega, \zeta) \hat{\tau}_3, \nonumber
\end{eqnarray}
where $g_i(x, \omega, \zeta)$ are scalar functions. For them  we derive three coupled linear differential equations:
\begin{eqnarray}
& &\zeta v_{F}\partial_x g_{1}(x) = - 2 \omega g_{2}(x)- 2 i \Delta_\zeta(x) g_3(x); \label{E1}\\
& &\zeta v_{F}\partial_x g_{2}(x) = - 2 \omega g_{1}(x);  \label{E2}\\
& &\zeta v_{F}\partial_x g_{3}(x) =  2 i \Delta_\zeta(x) g_1(x) \label{E3}
\end{eqnarray}  
(for brevity, only the coordinate dependence of $g_i$ is indicated above, but $\omega$ and $\zeta$ dependences are assumed). The normalization condition now reads 
\begin{eqnarray}
g_1^2(x)-g_2^2(x)+g_3^2(x) \equiv 1
 \label{norm}
\end{eqnarray} 
Note that Eqs.~(\ref{E1}--\ref{E3}) preserve normalization of $\hat{g}$ and so, it is sufficient to normalize it at one (arbitrary) point, $x$.

The $\hat{\tau}_1$ component of $ \hat{g}_{\omega}$ may be induced by an inhomogeneity of the gap (through the derivatives of $g_2$ and $g_3$), and disappears in the limit $\Delta \rightarrow {\rm const}$ or $\Delta \rightarrow 0$, {\it i.e.} in a bulk superconductor or a metal.  

To calculate the DOS using $ \hat{g}_{\omega}$ we need to analytically continue to real energies, $\epsilon$, via the substitution $\omega\rightarrow -i\epsilon + \delta$, where $\delta$ is a small positive constant. The general expression for DOS (here and below we take the temperature to be zero, $T=0$) is
\begin{eqnarray}
\frac{N(\epsilon, x)}{N_0}  = \frac{1}{4} \sum\limits_{\zeta = \pm} {\rm Re}\, [{\rm Tr}\,\{\hat{\tau}_3 \hat{g}( x, \epsilon,\zeta)\}].
\label{DOS1}
\end{eqnarray}
{\em I.e.}, it is proportional to the normal part of the quasiclassical Green's function.

Note that strictly speaking, $\hat{\Delta}$ should be self-consistently obtained within the theory. This, however, is unnecessary if we assume that the superconductivity in the wire at $x<0$ is induced by contact with a bulk superconductor (this likely represents the experimentally relevant situation in similar setups). Hence, we will treat the pair potential at $x<0$ as  an externally-imposed superconducting field. Also, in this geometry,  we can consider the normal and superconducting segments as parts of the {\it same} nanowire, since the variation in $\hat{\Delta}$ is externally-driven, rather then induced by a change in the physical properties of the wire. 
Let us add here that similar methods have been  used for the studies of superconducting nanowires in Refs.~\onlinecite{Neven,Abay}.

\section{Infinite wire}
\label{IW}

 First, we consider an infinite wire, with a superconducting pair potential that vanishes for positive $x$:  
$$
\hat{\Delta}\equiv \Delta_\zeta(x) \hat{\tau}_2 = \zeta \Delta_0 \theta(-x) \hat{\tau}_2,
$$ 
where $\Delta_0$ is a constant, and $\theta(x)$ is the step function. Note that $\hat{g}$ remains continuous at $x=0$, despite the discontinuity of $\hat{\Delta}$ there. 

Let us start from the superconducting half of the wire ($x<0$). Solving for $g_{1}(x)$ we get:
\begin{eqnarray}
 g_{1}(x,\omega,\zeta) = A_-(\omega,\zeta)  e^{\frac{2 \Omega}{v_F}x} 
 \label{g1}
\end{eqnarray} 
where $\Omega^2=\omega^2 + \Delta_0^2$, and the coefficient, $A_-(\omega,\zeta)$, is a function of frequency and, in principle, parameter, $\zeta$, to be determined by the boundary conditions. For obvious reasons, we have kept only the solution that is exponentially suppressed in the limit $x\rightarrow - \infty$.

Using  Eq. (\ref {g1}) and the Eilenberger equations~(\ref{E2}) and (\ref{E3}), we obtain the other components of the quasiclassical Green function for $x<0$ as follows
\begin{eqnarray}
& & g_{2}(x,\omega,\zeta)  = - i \frac{\Delta_\zeta}{\Omega} -  \zeta A_-(\omega,\zeta) \frac{\omega}{\Omega} e^{\frac{2 \Omega}{v_F}x}, \label{s2-} \\ 
& & g_{3}(x,\omega,\zeta)  = \frac{\omega}{\Omega} + i \zeta A_-(\omega,\zeta) \frac{\Delta_\zeta}{\Omega} e^{\frac{2 \Omega}{v_F}x},  \label{s3-} 
\label{g2g3}
\end{eqnarray} 
where we have required that the solution above reproduces the mean-field result for a uniform superconductor in the limit, $x\to-\infty$. Note that by doing so, we have automatically enforced normalization condition (\ref{norm}) for any $x$.

On the other side ($x>0$), the general solution is even  simpler, since $\Delta\equiv 0$. The spatial derivative of $g_{3}(x)$ vanishes, and we have to set it equal to the value in a bulk normal metal without any superconducting correlations -- $g_{3}(x) \equiv {\rm sgn}\,(\omega)$. By solving the equations we find for $x>0$:
\begin{eqnarray}
& & g_{1}(x,\omega,\zeta)  = A_+(\omega,\zeta) e^{-\frac{2 |\omega|}{v_F}x}, \label{s1+} \\ 
& & g_{2}(x,\omega,\zeta)  = \zeta {\rm sgn}\,(\omega) A_+(\omega,\zeta) e^{-\frac{2 |\omega|}{v_F}x}, \label{s2+} \\
& & g_{3}(x,\omega,\zeta)  =  {\rm sgn}\,(\omega) \label{s3+}. 
\end{eqnarray}
\begin{figure}[h]
\begin{center}$
\begin{array}{cc}
\includegraphics[width=0.5\textwidth]{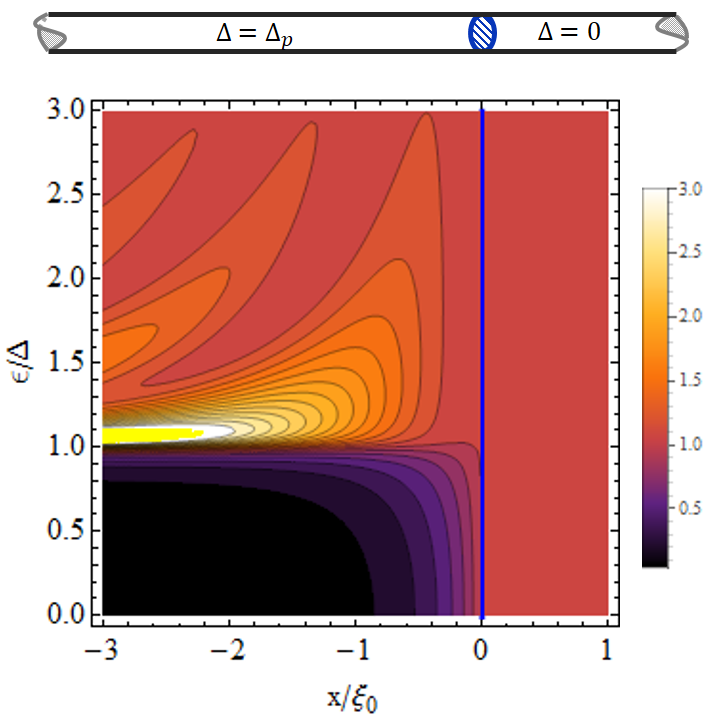}
\end{array}$
\end{center}
\caption{DOS of an infinite wire. The blue vertical line divides the superconducting and the normal parts of the system. We see that the DOS of the normal segment is everywhere equal to $1$. Calculations have been done with $\delta=0.03\Delta$. The solid yellow marks the regions that are beyond the plot range (where $N/N_0>3$).}
\label{Fig1}
\end{figure}
Finally, matching the solutions at $x<0$ with those at $x>0$, we find 
$$
A_-(\omega,\zeta)= A_+(\omega,\zeta)= \frac{i}{\Delta_0}\left[\omega- \Omega\, {\rm sgn}\,(\omega)\right],
$$
which, together with Eqs.~(\ref{g1}), (\ref{s2-}), and (\ref{s3-}) and (\ref{s1+}), (\ref{s2+}), and (\ref{s3+}), completely determines the solution on both sides. Note that $A_-$ is an even-in-momentum (no $\zeta$ dependence) and odd-in-frequency quantity. 

The functions $g_{1}$ and $g_{2}$ describe  superconducting correlations induced by the $p$-wave superconductor inside the normal part of the wire. Since $g_{1}(x,-\omega,\zeta) = - g_{1}(x,\omega,\zeta)$ and 
$g_{2}(x,\omega,-\zeta) = - g_{2}(x,\omega,\zeta)$, the former describes an odd-frequency $s$-wave component, while the latter an even-frequency $p$-wave component of the condensate wave-function. The fact that we obtain an odd-frequency component is not surprising -- the spacial inhomogeneity of the gap induces an $s$-wave-like  particle-particle correlations  proportional to $\hat{\tau}_1$ (it is similar to the $s$-wave-like pairing induced close to a vortex in a two-dimensional $p$-wave superconductor\cite{Tanuma}). The only way this induced pairing can satisfy the fermionic antisymmetry properties is for the correlations to vanish at equal times, {\em i.e.}, be odd in frequency\cite{OFP}. Indeed, it can be shown that the odd-frequency pairing correlations are ubiquitous close to boundaries and inhomogeneities of a superconductor\cite{Tanaka1} (these results can be straightforwardly extended to fermions with spin. In the case of a spin-singlet $s$-wave form of $\hat{\Delta}$ the inhomogeneity induces an odd-frequency $p$-wave pairing\cite{Buzdin,Efetov}).

 It is interesting 
to note that for the infinite wire, both odd-frequency and $p$-wave components not only appear on an equal footing, but in fact their amplitudes are identically equal to each other for any $x>0$
\begin{equation}
\label{g1=g2}
g_1^2(x,\omega,\zeta) \equiv g_2^2 (x,\omega,\zeta), \mbox{ for } x>0.
\end{equation}
This is despite the fact, that the presence of the odd-frequency part is a purely interfacial  effect, whereas the $p$-wave component ``leaks out'' from the interior  of the mean-field superconductor.

The normalization condition~(\ref{norm}) and Eq.~(\ref{g1=g2}) immediately yield the following result for the normal component of the quasiclassical Green's function, $g_{3}^2(x,\omega,\zeta) \equiv 1$ for $ x>0$ (which indeed is consistent with our explicit solution above -- see, Eq.~(\ref{s3+})). This implies that remarkably the normal component takes a constant value identical to that in a normal metal even in the presence of the proximity effect. Hence, in this geometry the DOS on the normal side is identical to that in a metal,  irrespective of the presence (or absence) of superconducting correlations  (see, Fig.~\ref{Fig1}).

To calculate the DOS on the other side, we analytically continue function (\ref{s3-}) to real energies. 
For $\epsilon<\Delta_0$ we get
\begin{eqnarray}
N(\epsilon, x)/N_0  = e^{2\sqrt{\Delta_0^2- \epsilon^2} x/v_f}, \mbox{ for } x<0
\label{fDOS1}
\end{eqnarray}
and, as discussed above,
\begin{eqnarray}
N(\epsilon, x)/N_0 \equiv 1, \mbox{ for } x>0.
\label{fDOS2}
\end{eqnarray}

From these results it is evident that the presence of the normal metal induces finite DOS for all energies in the superconductor, but this ``inverse proximity effect'' is exponentially suppressed away from the boundary (see, Fig. \ref{Fig1}). At the same time, the DOS of the normal half of the wire remains constant at its metallic value, as if there were no superconducting part attached to it (for the spin-singlet $s$-wave case, similar results have been obtained earlier \cite{Efetov, Buzdin}). This is despite the fact that the superconducting correlations are, in fact, present in the normal part.  This counter-intuitive result hinges on the precise balance between the odd-frequency and $p$-wave components (\ref{g1=g2}) and is in  fact accidental. The presence of a boundary or an impurity breaks it and immediately leads to a strong reconstruction of the metallic DOS into a more superconductor-like form, as discussed in the following sections.


\section{Semi-infinite wire with a finite normal section}
\label{IWFN}
Now we consider a semi-infinite wire with a finite metallic section, extending from $x=0$ to $x=L_M$. At its end, we assume a perfect specular reflection with the boundary condition: $\hat{g}( L_M, \omega, \zeta)=\hat{g}( L_M, \omega, -\zeta)$. This  mandates the vanishing of the $p$-wave component at the boundary: $g_2(L_M)=0$. With this condition in mind, we can write the solution of the Eilenberger equations -- Eqs. (\ref{E1}), (\ref{E2}), and (\ref{E3}) -- in the normal part ($0 < x < L_M$) as:
\begin{eqnarray}
& & g_{1}(x,\omega,\zeta)   = F(\omega) \cosh{\left[2 |\omega|(L_M-x)/v_F\right]}, \label{g1N}\\
& & g_{2}(x,\omega,\zeta)   = \zeta {\rm sgn}\,(\omega) F(\omega) \sinh{\left[2 |\omega|(L-x)/v_F\right]},   \label{g2N}\\
& & g_{3}(x,\omega,\zeta) = G(\omega). \label{g3N} 
\end{eqnarray}

\begin{figure}[h]
\begin{center}$
\begin{array}{cc}
\includegraphics[width=0.5\textwidth]{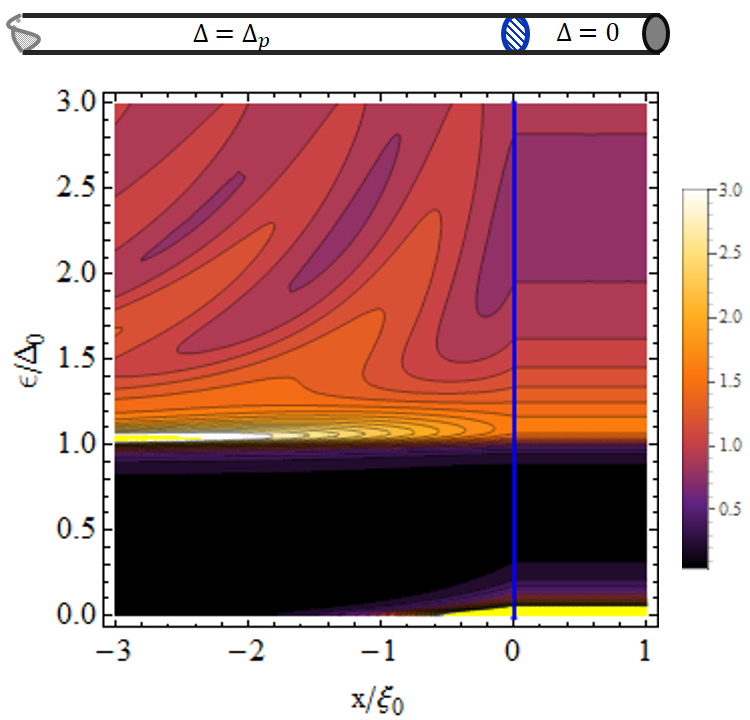}
\end{array}$
\end{center}
\caption{Plot of DOS of a semi-infinite wire with a finite normal segment of length $L_M=\xi_0$ (where $\xi_0 = 2v_F/(\pi \Delta_0)$ is the superconducting coherence length; see, text). There is an energy gap in the normal part of the system, and a delocalized zero-energy state inside it. The DOS of the normal part is constant in $x$. The conventions used are the same as those used for Fig. \ref{Fig1}. }
\label{Fig2}
\end{figure}

In the superconducting part the structure of $\hat{g}$ as given by Eqs. (\ref{g1}) -- (\ref{s3-}) is unchanged, but the coefficient $A_-$ is different. Matching the functions at $x=0$ we can determine the coefficients $A_-$, $F$ and $G$. From the continuity of $g_1$, it follows  that $A_-(\omega)=F(\omega) \cosh{\left(2 |\omega|L_M/v_F\right)}$. Matching of the $\hat{\tau}_2$-component gives:
\begin{eqnarray}
F(\omega)  = - \frac{i \Delta_0}{\tilde{\Omega} \sinh{\left(\tau_{M} |\omega| \right)}+ \omega \cosh{\left(\tau_{M} |\omega|\right)}},  \nonumber 
\end{eqnarray}
with $\tilde{\Omega} = {\rm sgn}\,(\omega) \Omega$ and $\tau_M = 2 L_M/v_F$. For exactly the same symmetry reasons as in the previous section, both $A_-(\omega)$ and $F(\omega)$ are odd-frequency quantities. Using the normalization condition (\ref{norm}) in the normal metal segment,  $G(\omega)^2=1-F^2(\omega)$,  we  get:
\begin{eqnarray}
G(\omega)  =  \frac{\Omega \cosh{\left(\tau_{M} |\omega|  \right)}+ |\omega| \sinh{\left(\tau_{M} |\omega| \right)}}{\tilde{\Omega} \sinh{\left(\tau_{M} |\omega| \right)}+ \omega \cosh{\left(\tau_{M} |\omega| \right)}}.  \nonumber 
\end{eqnarray}
Note that the denominator of $F(\omega)$ and $G(\omega)$  vanishes in the limit $\omega\rightarrow 0$.
\begin{figure}[h]
\begin{center}$
\begin{array}{cc}
\includegraphics[width=0.5\textwidth]{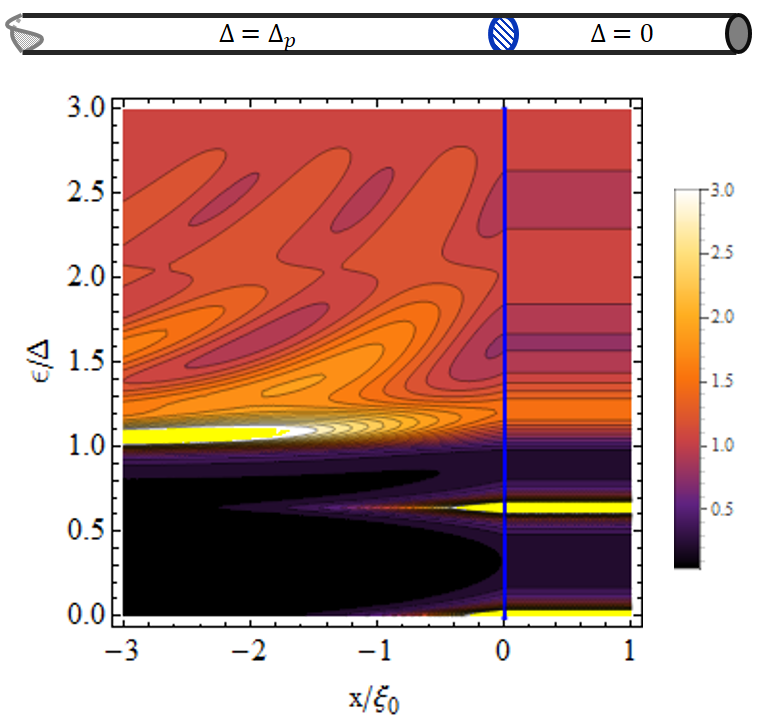}
\end{array}$
\end{center}
\caption{The same as on Fig. \ref{Fig2} but with $L_M=3 \xi_0$. We see that the longer normal part allows another in-gap bound state at finite energy.}
\label{Fig3}
\end{figure}

\begin{figure}[h]
 \begin{center}$
 \begin{array}{cc}
 \includegraphics[width=0.5\textwidth]{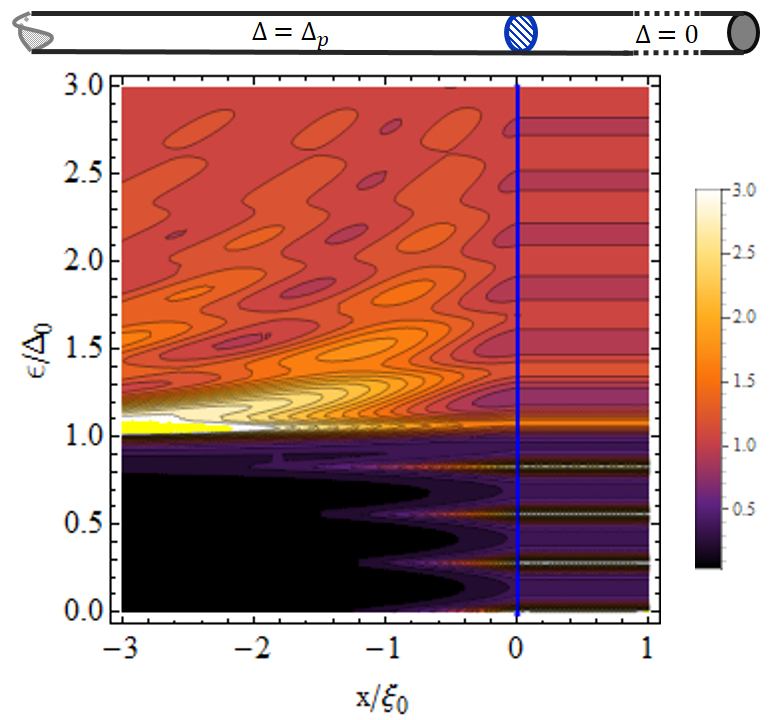}
 \end{array}$
 \end{center}
 \caption{The same as on Fig. \ref{Fig2} but with $L_M=8 \xi_0$. There are several bound states inside the gap. With the increase of $L_M$ they start to merge and the gap becomes ill-defined.}
 \label{Fig4}
 \end{figure}

 We can again obtain DOS from $g_3$ after an analytic continuation; see Eq.~(\ref{DOS1}). In contrast with the previous section, the DOS of the normal part of the wire is now strongly affected by  superconductivity. The proximity effect at the interface, combined with the reflection from the edge, lead to  a gap in the DOS for energies up to $\epsilon=\Delta_0$. Inside this gap, however, there are resonances, which demonstrate the presence of discreet Andreev bound states (ABS). They are similar to the  Caroli-de Gennes-Matricon energy states inside the cores of vortices. Their energies are determined by the condition that there imaginary part of the denominator of $G(\epsilon)$ vanishes, given by:
\begin{equation}
\tan \left( \frac{2 L_M}{v_F}\epsilon \right) = \frac{\epsilon}{\sqrt{\Delta_0^2 - \epsilon^2}},  \nonumber 
\end{equation}
(for $\epsilon<\Delta_0$).
As we see, the number and the positions of the ABS depend on the length of the normal section. A long metallic part admits many in-gap states -- compare Fig. \ref{Fig2}, Fig. \ref{Fig3} and Fig. \ref{Fig4}. There is an important exception, however -- the zero energy  state is always present, irrespective of the value of $L_M$.  This zero-energy contribution to DOS can be written in the form 
\begin{equation}
\label{Maj_res}
 N(\epsilon)/N_0 = \frac{\Delta_0}{4 L_M/\xi_0 + \pi}\, \delta(\epsilon),
\end{equation}
where $\xi_0=2 v_F/(\pi \Delta_0)$ is the superconducting coherence length. The weight of the peak is $L_M$-dependent, whereas its position is not. Note that in the limit of $L_M\rightarrow 0$  \cite{Honerkamp, Matsumoto1, Matsumoto2} this zero-energy feature squeezes into a ``usual'' localized Majorana fermion at the superconductor's end. 

As emphasized in Ref. \onlinecite{Tanaka2}, the delocalized zero-energy state (as well as all other in-gap states at finite energy) originate from the odd-frequency part of the quasiclassical Green's function in the Eilenberger formalism. As we saw in the previous section, however, an odd-frequency component can be present without a pronounced bound state. Thus  the presence of this component is, by itself,  not a sufficient condition for the appearance of a bound state (Majorana-state in particular). However, in the language of the quasiclassical Eilenberger theory, any Majorana-type state, both delocalized and localized (see also Sec.\ref{FW}), does originate from an odd-frequency component of the quasiclassical Green's function. 

 In the limit of a long normal part, the Andreev states begin to merge and eventually form a continuum with $N(\epsilon)/N_0=1$, and we recover the infinite wire case (shown on Fig. \ref{Fig1}). It is important to note that for any value of $L_M$ the bound states are uniformly spread over the normal part of the wire. This is consistent with the analytic\cite{JD1,Klinovaja} and numerical\cite{JD2,Chevallier} results obtained from the BdG equations. On the superconducting side the bound states are exponentially suppressed in $x$, with a decay length proportional to $\xi_0$.
   
\section{ Wire with an impurity}
\label{WI} 

We extend the previous case by considering a wire with a single impurity inside the finite normal part\cite{JD3}. We model the impurity by a delta-function-like scatterer  at the point $x=D$ (with $0<D<L_M$). We can write a general solution for the quasiclassical Green's function, $\hat{g}(x)$, for $x\in [0, D]$ as follows:
  \begin{eqnarray}
 \label{g1D}
 & &  g_{1}(x)  = F_L \Bigl[  \cosh{\left(\tau_x |\omega|\right)}\cosh{\left(\tau_{D} |\omega| \right)}   \nonumber \\
   & &\ \ \ \ \ \ \ \ \ \ \ \ + \alpha \sinh{\left(\tau_x |\omega|\right)}\sinh{\left(\tau_{D} |\omega|\right)} \Bigr],\\
  & &  g_{2}(x)  = \zeta {\rm sgn}\,(\omega) F_L \Bigr[ \sinh{\left(\tau_x |\omega|\right)}\cosh{\left(\tau_{D} |\omega|\right)}  \nonumber\\
  & &\ \ \ \ \ \ \ \ \ \ \ \  +  \alpha \cosh{\left(\tau_x |\omega|\right)}\sinh{\left(\tau_{D} |\omega|\right)} \Bigr], \\
  \label{g2D}
 & & \ \ \  g_{3}(x)= G_L,
 \label{g3D}
  \end{eqnarray}
  where we have introduced $\tau_x=2 (D-x)/v_F$  and $\tau_D=2 (L_M-D)/v_F$ (the subscript ``L" is for ``left"). $\alpha$ parameterizes the effect of the scatterer at $x=D$: $\alpha=0$ describes total reflection, whereas $\alpha=1$ total transmission. In both cases, we  recover the results from the previous section, but for different lengths of the normal part -- ($D$ or $L$ for $\alpha$ being $0$ or $1$ respectively). 
  
\begin{figure}[h]
   \begin{center}$
   \begin{array}{cc}
   \includegraphics[width=0.48\textwidth]{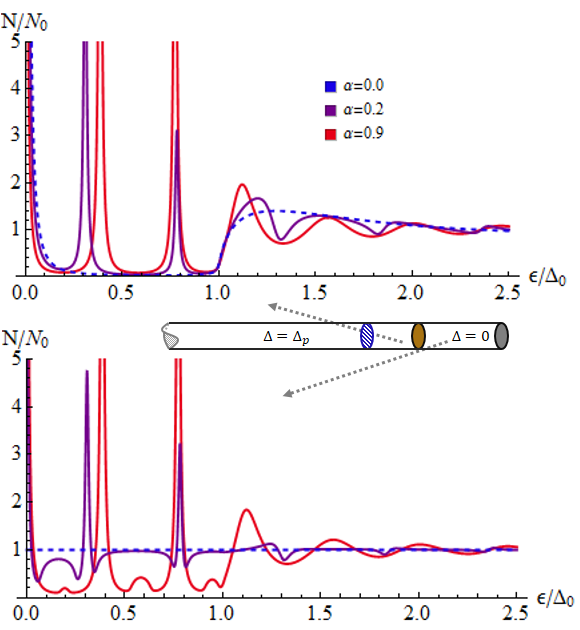}
   \end{array}$
   \end{center}
   \caption{DOS inside the normal part of the wire on the left (upper pane) and on the right (lower panel) of the impurity for $D= \xi_0$, $L_M= 6 \xi_0$, and for three different values of $\alpha$.}
   \label{Fig5}
   \end{figure} 
   
To obtain $F_L(\omega)$ and $G_L(\omega)$, we again have to match the functions with the ones on the superconducting side -- see Eqs.~(\ref{g1}), (\ref{s2-}), and (\ref{s3-}) -- at $x=0$. This yields $A_-(\omega)=F_L(\omega) \left\lbrace \cosh\left(\tau_0 |\omega|\right)\cosh\left(\tau_{D} |\omega|\right)+ \alpha \sinh\left(\tau_0 |\omega|\right)\sinh\left(\tau_{D} |\omega|\right) \right\rbrace$ and
 \begin{widetext}
 \begin{eqnarray}
F_L  =  \frac{-i \Delta_0}{\tilde{\Omega} \left[\sinh\left(\tau_0 |\omega|\right) \cosh\left(\tau_{D} |\omega|\right)+ \alpha \cosh\left(\tau_0 |\omega| \right)\sinh\left(\tau_{D} |\omega|\right)\right]+ \omega \left[ \cosh\left(\tau_0 |\omega|\right)\cosh\left(\tau_{D} |\omega|\right)+ \alpha \sinh\left(\tau_0 |\omega|\right)\sinh\left(\tau_{D} |\omega|\right)\right]},  \nonumber 
  \end{eqnarray}
  \end{widetext}
  and using the normalization condition  we can obtain: $G_L^2 = 1- \left[\cosh^2\left(\tau_{D} |\omega|\right)-\alpha^2 \sinh^2\left(\tau_{D} |\omega|\right) \right]F_L^2$.

For $x\in [D, L]$ we use Eqs. (\ref{g1N}) -- (\ref{g3N}), but with $F$ and $G$ replaced by $F_R$ and $G_R$ ($R$ for ``right"). By matching $g_{2}(D)$ (which has to be continuous) we get $F_R = \alpha F_L$ and from the normalization $G_R^2 = 1-F_R^2$.

The expressions for $g_i$ can also be obtained by treating the impurity as an effective boundary and using Zaitsev's boundary conditions\cite{Zaitsev, Kiesel}.They are conveniently written in terms of the symmetric and  antisymmetric components of $\hat{g}$: 
  \begin{equation}
  \hat{g}_{s,a}(x, \zeta)= \frac{1}{2}\left[ \hat{g}(x, \zeta) \pm \hat{g}(x,- \zeta)\right]
  \end{equation}
  The condition for the antisymmetric component,
  \begin{equation}
  \hat{g}_{a}(D-0, \zeta)= \hat{g}_{a}(D+0, \zeta),
  \end{equation}
  means that the $p$-wave component  $g_2$ has to be continuous at $x=D$ even for a nonzero $\alpha$. The symmetric component combining $g_1$ and $g_3$, however, is discontinuous there, and obeys the following matching condition:
  \begin{equation}
  -\alpha \left[\hat{g}_{s}(D+0)(1-\hat{g}_{a}(D)), \hat{g}_{s}(D-0)\right] = \hat{g}_{a}(D)\hat{g}_{s}^2(D+0). \nonumber
  \end{equation}
  Note that the conventional reflectivity is given by  
  $$
  R=\frac{1-\alpha}{1+\alpha}.
  $$

DOS is easy to obtain from  $G_L(\omega)$ and  $G_R(\omega)$. We show plots of DOS on both left and right side of the impurity and for  various values of $\alpha$ in Fig. \ref{Fig5}. For $\alpha=0$, the two parts of the normal segment are completely decoupled, and there are no superconducting correlations on the right of the impurity. The increase of $\alpha$ leads to ``leakage" of superconductivity to the right of the impurity and signatures of Andreev states at certain energies and gaps between them develop. On the left side the effect of the scatterer is to move the position the existing ABS and to induce new ones, if allowed by the total length of the normal section.  

Of particular interest here is the fate of the zero-energy resonance in the density of states. From the solution above, we obtain simple analytical expressions for the weight of this feature as a function of the relevant lengths and the coefficient, $\alpha$, characterizing scattering properties of the impurity. On the left of the impurity, we can write it as
\begin{eqnarray}
N(\epsilon)/N_0  = \frac{\Delta_0}{4 \left[D-\alpha (D -  L_M)\right]/\xi_0 + \pi}\, \delta(\epsilon),
\label{fDOS3}
\end{eqnarray}
and on the right we have
\begin{eqnarray}
N(\epsilon)/N_0  = \frac{\alpha \Delta_0}{4 \left[D-\alpha (D - L_M)\right]/\xi_0 + \pi}\, \delta(\epsilon).
\label{fDOS4}
\end{eqnarray}
We see that for any finite $\alpha$ the zero-energy peak is spread over the entire normal segment. However, its weight on the two sides of the impurity can differ significantly, especially for small $\alpha$ (this is consequence of the discontinuity of $g_3$ at $x=D$). Thus, the presence of strong scatterer cannot completely localize the zero-energy (Majorana) mode in  part of the normal wire, but it can lead to uneven distribution of its weight. 

In the regime $D<\xi_0\ll L_M$ our results are qualitatively consistent with Ref.~[\onlinecite{Gibertini}] (see Fig. 5 there).

\section{Finite wire}
\label{FW}
We now turn to the case of a finite wire with the superconducting part from $-L_p$ to $0$, and the normal part from $0$ to $L_M$. We are assuming perfect reflection on both external boundaries. 

As a sanity check, we expect the solution on the superconducting side to contain a pronounced zero-energy state associated with a Majorana fermion  localized near the superconductor's boundary at $x=-L_p$. This is well-known from the elementary solution of the Kitaev chain to which the section at $x<0$ is equivalent (in the limit of $L_p \gg \xi$), but now we look for manifestations of the Majorana fermion in the different language of the quasiclassical Green function (which contains less information than the BdG theory since short length-scales have been integrated out).

Turning to the technical details, we can write a general solution,  $g_i$, to the Eilenberger equation in the following form:  
\begin{eqnarray}
& & g_{1}(x) = A e^{\frac{2 \Omega}{v_F}x} + B e^{-\frac{2 \Omega}{v_F}(x+L_p)};\\
& & g_{2}(x) = \zeta  \frac{\omega}{\Omega} \left[ - A e^{\frac{2 \Omega}{v_F}x} + B e^{-\frac{2 \Omega}{v_F}(x+L_p)}  - C\right]; \\
& & g_{3}(x) = \frac{i \Delta_0}{\Omega} \left[  A e^{\frac{2 \Omega}{v_F}x} - B e^{-\frac{2 \Omega}{v_F}(x+L_p)}  - \frac{\omega^2}{\Delta^2} C\right]. 
\end{eqnarray}
Note that the coefficients $A$, $ B $, and $ C $ depend on $\omega$ and, in principle, $\zeta$, but we omit these dependencies for the sake of brevity. These functions are constrained by the boundary condition $g_2(-L_p)=0$ and the normalization.
The former requires $C = B - A e^{-\frac{2 \Omega}{v_F} L_p}$,
and writing the latter at $ x=-L_p $ gives
\begin{eqnarray}
4 A B e^{-\frac{2 \Omega}{v_F} L_p} - \omega^2 \left(B-A e^{-\frac{2 \Omega}{v_F} L_p}\right)^2=1. \nonumber
\label{NormCondLM}
\end{eqnarray}
\begin{figure}[h]
\begin{center}$
\begin{array}{cc}
\includegraphics[width=0.48\textwidth]{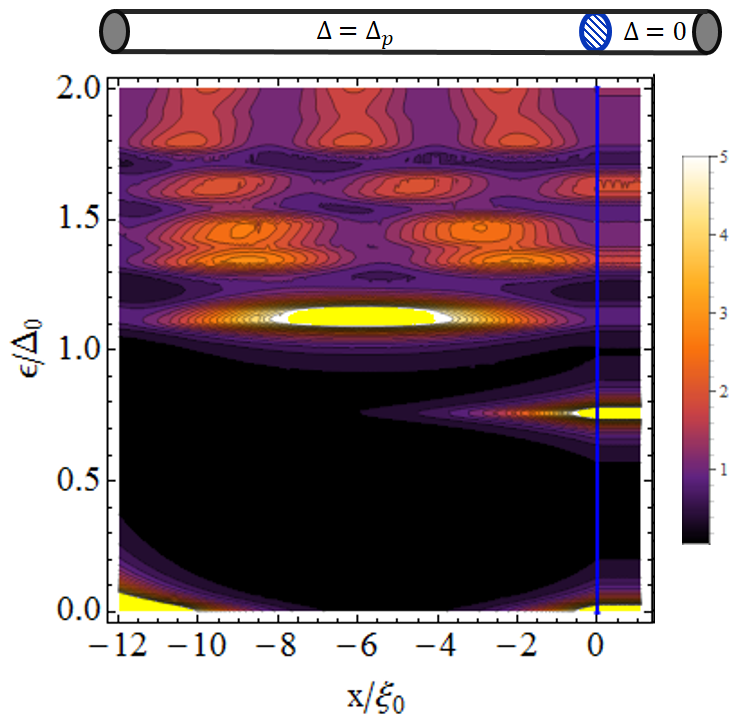}
\end{array}$
\end{center}
\caption{DOS of a finite wire with length of the superconducting and normal segments $L_p= 12 \xi_0$ and  $L_M= 3 \xi_0$ respectively. At the left edge there is a single localized zero-energy state inside the gap. The bound states at the right edge are spread over the entire normal segment.}
\label{Fig6}
\end{figure} 
For the normal part of the wire, Eqs. (\ref{g1N})-(\ref{g3N}) are still valid (with $F$ and $G$ appropriately modified). Matching $ g_1$ and $g_2$ at $ x=0$, we get:
\begin{eqnarray}
& & F \cosh \left(\tau_M |\omega| \right)=A e^{-\frac{2 \Omega}{v_F}L_p} + B;  \,\nonumber\\
& & F \sinh \left(\tau_M |\omega| \right) = \frac{\omega}{\Omega} (A + B)  \left( e^{-\frac{2 \Omega}{v_F}L_p} - 1   \right), \nonumber
\end{eqnarray}
with $\tau_M =2L_M/v_f$, as defined in Sec.~\ref{IWFN}. Combining these equations, we get
\begin{eqnarray}
A = \frac{\beta \Delta }{\sqrt{-\omega^2 + \beta e^{-\frac{2 \Omega}{v_F}L_p}\left(4 \Delta^2 + 2 \omega^2 - \omega^2 e^{-\frac{2 \Omega}{v_F}L_p}\right)}},\nonumber
\end{eqnarray} 
and $ A= \beta B$, where $\beta$ is defined as follows:
\begin{eqnarray}
\beta = \frac{\omega \cosh \left(\tau_M |\omega| \right) \left( e^{-\frac{2 \Omega}{v_F}L_p} - 1   \right)- \Omega \sinh \left(\tau_M |\omega| \right) e^{-\frac{2 \Omega}{v_F}L_p}  }{\Omega \sinh \left(\tau_M |\omega| \right)-\omega \cosh \left(\tau_M |\omega| \right) \left(e^{-\frac{2 \Omega}{v_F}L_p} - 1   \right)}.\nonumber
\end{eqnarray}
In addition we have  $C = (\beta - e^{-\frac{2 \Omega}{v_F} L_p}) B $, and obtaining $F$ and $G$ is also straightforward. 

We can now study the DOS along the finite wire. As in the case of the semi-infinite wire, there are in-gap states, uniformly distributed over the entire normal part. Their number and energies depend on $L_M$, which in essence introduces finite-size quantization effect similar to those in elementary single-particle quantum mechanics. On the other edge ($x=-L_p$) the most prominent feature is the single bound state at zero energy and localized in $x$ (see, Fig. \ref{Fig6}).  Its contribution to DOS for $L_p \gg \xi_0 $ is 
\begin{equation}
\label{Maj_res2}
 N(\epsilon, x)/N_0 = \frac{\Delta_0}{\pi}\, e^{-\frac{4 }{\pi \xi_0}(x + L_p )}\, \delta(\epsilon),
\end{equation}
with $x$ close to $-L_p$. We see the characteristic exponential decay away from the edge. This is, of course, a feature associated with the ``usual'' localized Majorana fermion.  

\begin{figure}[h]
 \begin{center}$
 \begin{array}{cc}
 \includegraphics[width=0.48\textwidth]{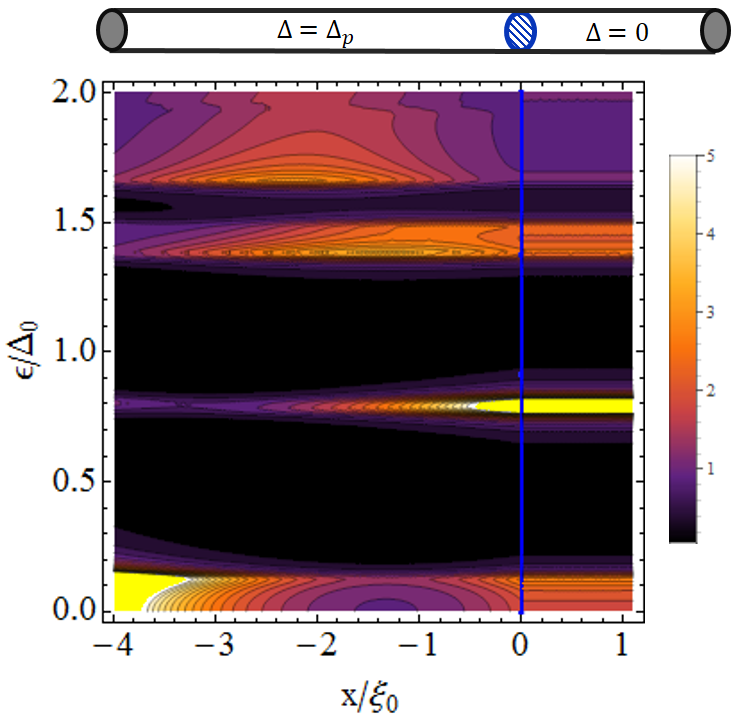}
 \end{array}$
 \end{center}
 \caption{The same as on Fig. \ref{Fig6} but with lengths $L_p= 4 \xi_0$ and  $L_M= 3 \xi_0$. We see the splitting of the zero-energy peak due to the mixing of the left and right edge modes.}
 \label{Fig7}
 \end{figure}

\begin{figure}[h]
 \begin{center}$
 \begin{array}{cc}
 \includegraphics[width=0.48\textwidth]{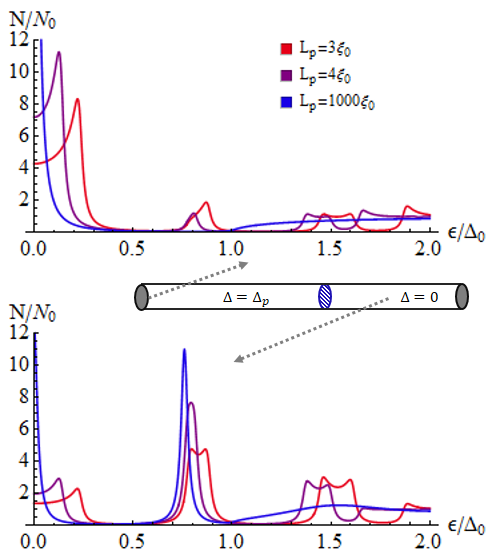}
 \end{array}$
 \end{center}
 \caption{DOS at the $x=-L_p$ point (upper panel), and inside the normal segment (lower panel) for three different values of $L_p$ (and $L_M$ fixed at $3 \xi_0$). Notice that all bound states show splitting, but the effect is the strongest for the Majorana-type state.}
 \label{Fig8}
 \end{figure}

In the limit of $L_p \gg \xi_0 $, we recover the results from the previous sections. However, as the superconducting part of the wire becomes shorter, the localized Majorana fermion and the delocalized zero-energy state in the normal part begin to mix. The two zero-energy DOS peaks now shift to small, but finite $\delta\epsilon$ (which is the same on both sides); this is shown on Fig. \ref{Fig7}  and Fig. \ref{Fig8}. We can obtain an analytic expression for this splitting by looking at the denominator of $A$. Expanding it in powers of $\epsilon$ and $\exp{(-4 L_p/\pi \xi_0)}$ we get 
 \begin{eqnarray}
 \delta \epsilon = \frac{2\Delta }{\sqrt{1+ \frac{4 L_M}{\pi \xi_0}}} \exp{\left(-\frac{4 L_p}{\pi \xi_0}\right)}.
 \label{split}
 \end{eqnarray} 
  
  We need to emphasize that care is required in interpreting and applying this result for the splitting. To understand the subtleties involved, recall that splitting of (localized) Majorana modes  has been considered before by one of the authors~\cite{VG2, VG3} and it was shown that it originates from the tunneling of Majorana fermions between their localized positions (e.g., in vortex cores or wire's ends) through the superconducting medium (which serves as a tunneling barrier in single-particle quantum mechanics). The corresponding  overlap of the two states determines the magnitude and sign of the splitting, and it is very sensitive to the microscopic details of the single-Majorana wave-functions involved, including the length-scales of order Fermi wave-length. This leads to a fast oscillating splitting ($\propto \cos(k_F r)$, which can be qualitatively interpreted as a ``square root of a Friedel oscillation''). These fast oscillations are superimposed on an exponential decay at larger length-scales of order coherence length, $\xi_0$. The quasi-classical Eilenberger formalism integrates out the smallest length-scales from the outset and is strictly  speaking valid only at length-scales, $k_F^{-1} \ll r$. Therefore, while it can correctly capture the exponential decay of the splitting, the ``Friedel oscillation'' effects are beyond the method's reach. In particular, Eq.~(\ref{split}) above gives the mean value of the spitting (qualitatively, it is the modulus of the splitting averaged over distances much larger than Fermi wave-length , but smaller than other length scales). However, the result -- Eq.~(\ref{split}) -- is not incorrect, but rather relates to a certain experimental setup, where geometry of either contacts  or structure's boundaries may enforce such an averaging (e.g., if the relevant length-scales  are much larger than Fermi wave-length). If this condition is not satisfied (e.g., the DOS is probed at a sharp wire's end with a tunneling tip), Eq.~(\ref{split}) gives only an estimate of the splitting. In either case, an important conclusion of this section is that localized and delocalized Majorana modes split in much the same way, as two localized modes do.
 
 It is also interesting to note that the finite energy Andreev states on the left also tend to split as $L_p$ decreases. 

\section{Discussion and conclusion}
\label{DC}

In this paper, we used the quasiclassical Eilenberger theory to study the topological proximity effects in superconducting nanowires with different geometries. One of the goals of the work has been to further explore the relationship between the phenomena of odd-frequency pairing and topological superconductivity, including Majorana fermion physics. To briefly summarize, it appears that the Majorana modes (both localized and delocalized) indeed originate from an odd-frequency component of the anomalous part of the quasiclassical Green function and show up in the Eilenberger formalism as a zero-frequency pole in the Green function. However, the mere existence of an odd-frequency pairing does not necessarily imply the existence of a Majorana fermion.

On the methodological side, we have demonstrated that the Eilenberger formalism is preferable to (the more general) Gor'kov equations and Bogoliubov-de~Gennes methods, whenever the Fermi wave-length is the smallest length-scale in the problem and one is interested in observables (in particular, the density of electronic states) at length scales much larger than that. Of course all methods yield identical results for the same observables, but the Eilenberger method is much simpler than the more microscopic approaches and often allows one to readily obtain analytic results, where the other approaches require elaborate calculations and numerical simulations. 

Our specific results include simple analytical solutions for the density of states in a model system, which involves a nanowire with a $p$-wave mean-field superconducting order parameter on one side and no such order parameter on the other. While this model may seem artificial, it  retains the main qualitative features of more realistic setups that involve a bulk superconductor proximity-inducing topological superconductivity in a part of a wire. 
The key results in our model and the latter models also share qualitative similarities. Returning to such specific results, we find, somewhat counterintuitively, that in a clean infinitely-long wire, the DOS in the normal part does not change at all compared to a metal despite the presence of strong superconducting correlations.  On the other hand, superconductivity on the superconducting side becomes ``almost'' gapless close to the interface (note the similarities with the ``inverse proximity effect" discussed in Ref. \onlinecite{Stanescu}). The former result hinges on an accidental cancellation of odd-frequency and $p$-wave pairing terms, but this balance is disturbed by  a slightest perturbation.

In particular, if the normal part has finite length, a gap is proximity induced there, and delocalized discreet states appear inside the gap, including a Majorana-type mode at zero energy. The weight of the corresponding zero-energy peak in the density of state scales with the length of the metallic segment in a simple fashion, see Eq.~(\ref{Maj_res}). This dependence does not appear to rely much on the details of the topological heterostructure and should be accessible in tunneling experiments with existing setups, where the effective length of the metallic segment may be modified by electrostatic means. Most importantly,  the Eilenberger theory makes it transparent that if a potential Majorana host is not in direct contact with a superconductor, it is impossible to induce a localized Majorana fermion there (which would have all other sought-after attributes such as quantized zero-bias tunneling anomaly near a wire's end and well-defined non-Abelian braiding statistics with respect to other such modes). This conclusion is based on a simple observation that the spatial dependence of any fermionic mode is proportional to the superconducting gap, which in turn can only appear due to intrinsic interaction or immediate proximity to a bulk superconductor, but is exactly zero in a non-interacting metal. 

We also have found that in a finite wire, the zero-energy peaks associated with a localized Majorana fermion in the superconducting part and delocalized mode in the normal part split up to the same finite energy. We provide an analytical formula -- see, Eq.~(\ref{split}) --  that estimates the mean value of the splitting (averaged over distance larger than the Fermi wave-length). This quasiclassical treatment of this phenomenon however illustrates some limitations of the Eilenberger approach, which misses length scales of order Fermi wave-length, where the splitting is expected to oscillate strongly.

\section*{ACKNOWLEDGMENTS}
We gratefully acknowledge useful discussions with Philip Brydon,  Jay Deep Sau and Sankar Das Sarma. This work was supported by DOE-BES (DESC0001911) and Simons Foundation.

\end{document}